\documentclass{article}
\usepackage{spconf,amsmath,amssymb,graphicx}
\usepackage{booktabs}
\usepackage{multirow}
\usepackage{dblfloatfix}
\usepackage{url}

\newcommand{\R}{\mathbb{R}}
\newcommand{\calL}{\mathcal{L}}
\newcommand{\low}{\mathrm{L}}
\newcommand{\high}{\mathrm{H}}
\newcommand{\up}{\mathrm{U}}
\newcommand{\down}{\mathrm{D}}

\title{FFMT: FREQUENCY-FACTORIZED MULTI-RATE TOKENIZATION FOR MUSIC-TO-DANCE GENERATION}

\name{
Qirong Guo$^{1*}$,
Ruijie Li$^{1*}$,
Zeyan Li$^{2}$,
Sicheng Shu$^{3}$,
Guanzhi Ding$^{1}$,
Weihang Ding$^{4}$
}

\address{
\small $^{1}$The Hong Kong University of Science and Technology (Guangzhou)\\
\small $^{2}$Shanghai Jiao Tong University\quad
$^{3}$Duke University\quad
$^{4}$University of California, Berkeley
}

\begin{document}
\ninept
\maketitle
\begingroup
\renewcommand{\thefootnote}{\fnsymbol{footnote}}
\footnotetext[1]{Equal contribution.}
\endgroup

\begin{abstract}
Dance motion combines slowly varying body configurations with rapid local movements. We investigate how explicit temporal-frequency decomposition can guide the temporal resolution and codebook capacity of discrete motion representations. We introduce frequency-factorized motion tokenization (FFMT), which separates root-relative motion into a low-frequency choreographic scaffold and a complementary high-frequency residual. FFMT uses a lower token rate and a smaller codebook for the low-frequency component, and a higher token rate and a larger codebook for the high-frequency residual. This representation further induces a plan-then-articulate autoregressive factorization: low-frequency tokens establish coarse choreography, while high-frequency tokens realize local motion details conditioned on the generated plan and music. Experiments on AIST++ and FineDance show strong distributional fidelity while preserving realistic diversity and music--motion synchronization. Ablations consistently support frequency factorization, asymmetric temporal allocation, and low-to-high conditioning.
\end{abstract}

\begin{keywords}
music-to-dance generation, motion tokenization, vector quantization, frequency decomposition, autoregressive generation
\end{keywords}

\section{Introduction}
Music-driven 3D dance generation aims to synthesize movements that follow musical structure while maintaining coherent choreography and expressive local articulation. Recent autoregressive and diffusion-based approaches have advanced music-conditioned motion synthesis~\cite{li2021aistplusplus,siyao2022bailando,tseng2023edge,li2024lodge,zhang2024badm,luo2024popdg}. For systems built on discrete motion tokens, the tokenizer defines the units over which the generator learns music--motion relationships~\cite{siyao2023bailandopp,yang2026tokendance,ghosh2025duetgen,li2025souldance}. Its temporal resolution and representational capacity shape how slowly varying body configurations and rapid local movements are encoded, making the organization of motion tokens an important consideration for capturing the temporal complexity of dance.

Prior work has explored temporal hierarchies in dance tokenization. DanceMeld~\cite{gao2023dancemeld} and DuetGen~\cite{ghosh2025duetgen} learn coarse and fine motion codes at different temporal resolutions, with DuetGen conditioning fine-level generation on the coarse tokens. Frequency information has also been incorporated into reconstruction and generation guidance: ExpressiveBailando~\cite{huang2024expressivebailando} applies Fourier-domain supervision to enhance motion decoding, while InfiniteDance~\cite{li2026infinitedance} constructs conditioning features by processing frequency bands of retrieved reference motions. Temporal resolution alone does not specify the frequency content assigned to each stream. We therefore investigate explicit motion-frequency components as quantization targets for allocating token rates and codebook capacity.

This leads to a \emph{temporal capacity allocation} problem. Token rate controls temporal resolution: a low rate may miss rapid changes, whereas a high rate may redundantly encode slowly varying motion. Codebook size determines the states available per token: a small codebook may merge distinct motion patterns, while a larger one provides more alternatives. The timing and amplitude of local movements must remain coordinated with evolving body configurations, making cross-component dependencies another consideration. The question is \emph{how to allocate temporal resolution and codebook capacity across frequency components while preserving their coordination during music-conditioned generation?}

We address this question with \textbf{frequency-factorized motion tokenization (FFMT)}. A fixed Gaussian filter separates root-relative motion into a low-frequency component and its complementary high-frequency residual before quantization. The low-frequency stream uses a lower token rate and a smaller codebook, while the high-frequency stream uses a higher rate and a larger codebook. A music-conditioned \emph{plan-then-articulate} generator first predicts low-frequency tokens, then generates high-frequency tokens conditioned on the low-frequency sequence and music.

Our contributions are threefold: (1) we formulate discrete dance representation as a temporal capacity allocation problem and develop a tokenizer with complementary frequency streams; (2) we allocate frequency-dependent token rates and codebook capacities and couple them with plan-then-articulate generation, where high-frequency dynamics are conditioned on generated low-frequency structure; and (3) we validate these design choices on AIST++ and FineDance through ablations of frequency factorization, multi-rate allocation, cross-frequency conditioning, and the decoding prior.

\section{Method}
\begin{figure}[!t]
\centering
\includegraphics[width=0.98\columnwidth]{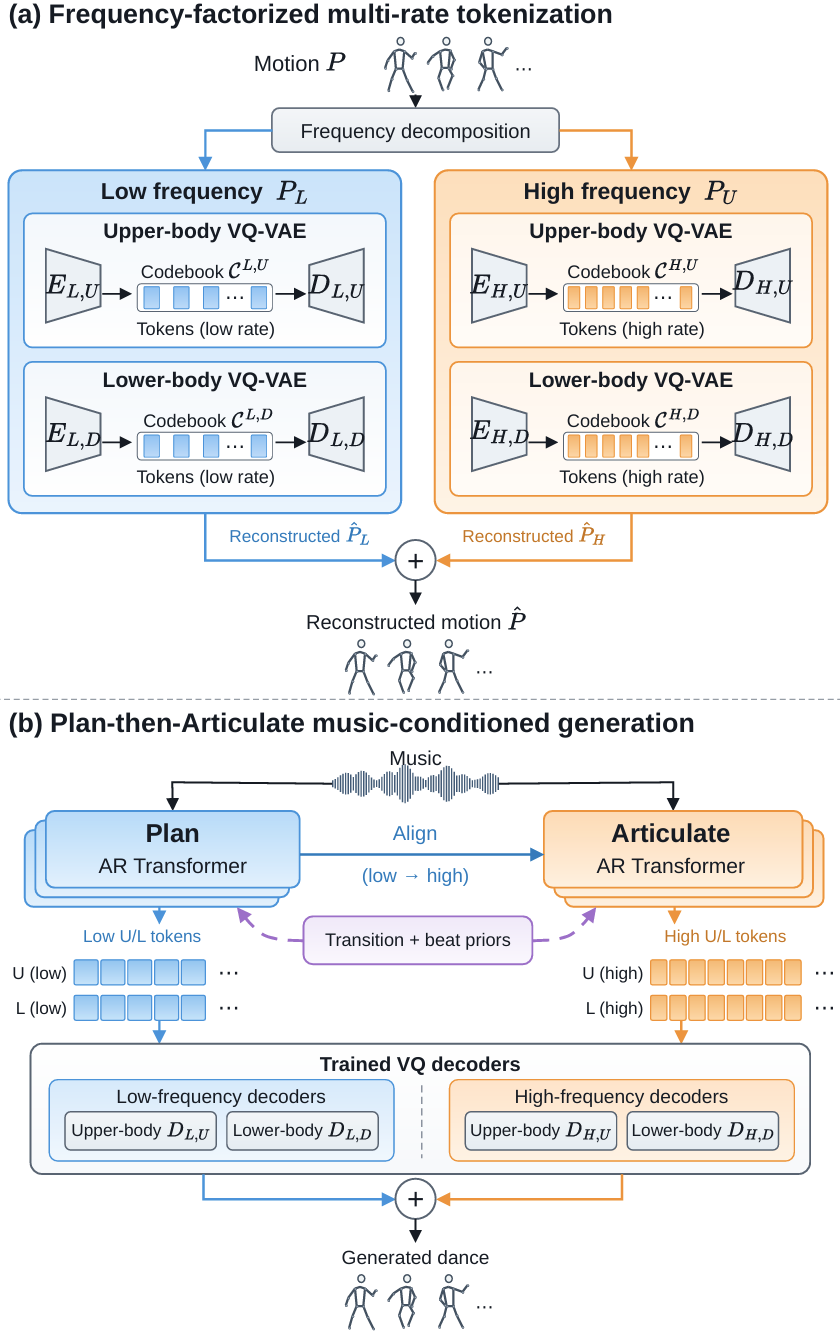}
\caption{Overview of FFMT.
(a) Frequency-factorized tokenization assigns different temporal
resolutions and codebook sizes to low-frequency structure and
high-frequency residuals, with separate upper/lower-body codebooks.
(b) Plan-then-articulate generation predicts low-frequency tokens
first and conditions high-frequency generation on the aligned
low-frequency context and music. The resulting tokens are decoded
into rotations and root motion, with a training-derived prior
used during inference.}
\label{fig:overview}
\end{figure}

\textbf{Overview.} Given music $\mathbf M$, FFMT generates a 3D dance sequence $\mathbf X$ through a VQ-VAE + autoregressive pipeline (Fig.~\ref{fig:overview}). The tokenizer produces low- and high-frequency motion codes, whose music-conditioned generation factorizes as:
\begin{equation}
p(\mathbf Z^{\low},\mathbf Z^{\high}\mid\mathbf M)
=p(\mathbf Z^{\low}\mid\mathbf M)
p(\mathbf Z^{\high}\mid\mathbf Z^{\low},\mathbf M).
\label{eq:factor}
\end{equation}
Each frequency stream uses upper/lower-body codebooks~\cite{siyao2022bailando}. The generated codes are decoded into joint rotations and root motion.

\subsection{Frequency-Factorized Motion Tokenization}
Let $\mathbf{P}\in\R^{T\times J\times 3}$ denote a sequence of $J$ 3D joints over $T$ frames. Before tokenization, we remove global root translation to obtain root-relative joint positions $\bar{\mathbf{P}}$.

We apply a fixed Gaussian low-pass filter along time. For radius $R$ and bandwidth $\sigma$, the normalized kernel is
\begin{equation}
g(\tau)=\frac{\exp(-\tau^2/2\sigma^2)}
{\sum_{u=-R}^{R}\exp(-u^2/2\sigma^2)},\quad \tau\in[-R,R].
\label{eq:gaussian}
\end{equation}
Using replicated boundary values, we define
\begin{equation}
\mathbf{P}^{\low}_{t}=\sum_{\tau=-R}^{R}g(\tau)\bar{\mathbf{P}}_{t+\tau},
\qquad
\mathbf{P}^{\high}=\bar{\mathbf{P}}-\mathbf{P}^{\low}.
\label{eq:decomp}
\end{equation}
Here $\mathbf{P}^{\low}_{t}\in\R^{J\times3}$ is the low-frequency component at frame $t$; stacking all $T$ frames yields $\mathbf{P}^{\low}\in\R^{T\times J\times3}$. Thus $\bar{\mathbf{P}}=\mathbf{P}^{\low}+\mathbf{P}^{\high}$ exactly. The high-frequency branch quantizes the \emph{complementary residual}. This additive construction preserves the complete position signal before quantization while exposing two temporal regimes for separate capacity allocation.

Each frequency component $\mathbf P^f$, $f\in\{\low,\high\}$, is partitioned along the joint dimension into body-part components $\mathbf P^{f,b}$, where $b\in\{\up,\down\}$ indexes the upper and lower body, respectively. The two components contain disjoint joint subsets and retain their original joint correspondences. Each of the four frequency--body components is modeled by a dedicated VQ-VAE branch with its own encoder, codebook, and position decoder.

Each branch $(f,b)$ uses an encoder $E_{f,b}$ with temporal stride $r_f$ frames. For a $T$-frame input, it produces $T_f=T/r_f$ latent vectors of dimension $d$, so $E_{f,b}(\mathbf P^{f,b})\in\R^{T_f\times d}$. Let $\mathcal{C}^{f,b}=\{\mathbf e^{f,b}_k\}_{k=1}^{K_f}$ denote its $K_f$-entry codebook, with $\mathbf e^{f,b}_k\in\R^d$. At token position $i=1,\ldots,T_f$, nearest-neighbor quantization gives
\begin{equation}
z^{f,b}_{i}=\arg\min_{k}\left\|
E_{f,b}(\mathbf{P}^{f,b})_{i}-\mathbf e^{f,b}_{k}
\right\|_2^2.
\label{eq:vq}
\end{equation}
FFMT uses $r_{\low}>r_{\high}$ and $K_{\low}<K_{\high}$. The larger low-frequency stride gives a lower token rate, paired with a smaller codebook; the high-frequency residual uses a higher token rate and a larger codebook.

\subsection{Motion Reconstruction and Tokenizer Training}
\noindent\textbf{Position reconstruction.} Let $D_{f,b}$ denote the position decoder associated with branch $(f,b)$. Its token sequence is reconstructed as
\begin{equation}
\hat{\mathbf P}^{f,b}=D_{f,b}(\mathbf Z^{f,b}),
\qquad f\in\{\low,\high\},\;b\in\{\up,\down\}.
\label{eq:posdecode}
\end{equation}
For each frequency stream, the decoded body-part components are reassembled in the original joint order to obtain $\hat{\mathbf P}^{f}$, yielding the full root-relative reconstruction $\hat{\mathbf P}=\hat{\mathbf P}^{\low}+\hat{\mathbf P}^{\high}$.

Let $\mathbf h_i^{f,b}=E_{f,b}(\mathbf P^{f,b})_i$ denote the encoder output at token position $i$. Each branch is trained with
\begin{align}
\calL_{\mathrm{branch}}^{f,b}={}&
\sum_{m=0}^{2}\left\|\Delta^m\mathbf P^{f,b}
-\Delta^m\hat{\mathbf P}^{f,b}\right\|_1 \nonumber\\
&+\frac{\beta}{T_fd}\sum_{i=1}^{T_f}
\left\|\mathbf h_i^{f,b}-\operatorname{sg}\!\left(
\mathbf e_{z_i^{f,b}}^{f,b}\right)\right\|_2^2 \nonumber\\
&+\frac{\eta}{T_fd}\sum_{i=1}^{T_f}
\left\|\mathbf h_i^{f,b}\right\|_2^2.
\label{eq:branchloss}
\end{align}
The position objective combines the average branch loss with full-motion reconstruction and temporal consistency:
\begin{align}
\calL_{\mathrm{pos}}={}&\frac{1}{4}\sum_{f,b}\calL_{\mathrm{branch}}^{f,b} \nonumber\\
&+\sum_{m=0}^{2}\lambda_m\left\|\Delta^m\bar{\mathbf P}
-\Delta^m\hat{\mathbf P}\right\|_1.
\label{eq:posloss}
\end{align}
All reconstruction and temporal terms are evaluated on the original frame grid after decoding. The branch losses supervise their fixed frequency--body targets, while the full-motion terms jointly optimize the branches through the recombined motion.

\noindent\textbf{Rotation/root decoding.} After position-space VQ training, we freeze the position encoders and codebooks and train the motion decoder~\cite{siyao2023bailandopp}. For each body part $b$, $D_F^{f,b}$ decodes quantized codes into frame-resolution features, which are fused as
\begin{equation}
\mathbf F^b=\Phi_b\!\left(
[D_F^{\low,b}(\mathbf Z^{\low,b});D_F^{\high,b}(\mathbf Z^{\high,b})]
\right),
\label{eq:fusion}
\end{equation}
where $[\cdot;\cdot]$ denotes channel concatenation and $\Phi_b$ is a temporal fusion network. The fused features predict joint rotations, with lower-body features additionally predicting root velocity. Global translation is obtained by integrating the predicted root velocity from a specified initial root position. The decoder is trained with rotation/root reconstruction and temporal consistency objectives.

\subsection{Plan-then-Articulate Music-Conditioned Generation}
After tokenizer training, each dance is represented by four streams $\{\mathbf Z^{\low,\up},\mathbf Z^{\low,\down},\mathbf Z^{\high,\up},\mathbf Z^{\high,\down}\}$. Let $c_t^f=(z_t^{f,\up},z_t^{f,\down})$ denote the upper/lower-body token pair at token timestep $t$ within frequency stream $f$. Low- and high-frequency token pairs are generated by separate autoregressive Transformers, each comprising stacked Transformer layers. The high-frequency model additionally conditions on the low-frequency tokens. Following Eq.~\eqref{eq:factor}, generation proceeds in two stages.

\noindent\textbf{Plan.} Because the motion streams use different token rates, music context is encoded at the corresponding temporal resolutions. A local-context music Transformer produces $\mathbf A^{\low}$ and $\mathbf A^{\high}$ by aggregating neighboring music frames and downsampling them to $T_{\low}$ and $T_{\high}$ steps. The low-frequency Transformer first autoregressively establishes coarse choreography:
\begin{equation}
p(\mathbf Z^{\low}\mid\mathbf M)
=\prod_{t=1}^{T_{\low}}
p(c_t^{\low}\mid
\mathbf Z_{<t}^{\low},\mathbf A^{\low}).
\label{eq:plan}
\end{equation}

\noindent\textbf{Articulate.} The high-frequency stage realizes residual dynamics conditioned on the generated plan. Since the two streams have different token rates, each fine timestep is aligned to the coarse sequence by
\begin{equation}
\pi(t)=1+\left\lfloor\frac{(t-1)T_{\low}}{T_{\high}}\right\rfloor,
\label{eq:align}
\end{equation}
where $t=1,\ldots,T_{\high}$ and $\pi(t)\in\{1,\ldots,T_{\low}\}$. With $T_{\high}=2T_{\low}$, fine timesteps $2i-1$ and $2i$ share coarse index $i$. A bounded neighborhood around $\pi(t)$ is embedded as low-frequency context $\mathbf C_t^{\low}$. This alignment selects the coarse tokens used for conditioning; it does not resample their discrete indices. The high-frequency distribution is
\begin{equation}
p(\mathbf Z^{\high}\mid\mathbf Z^{\low},\mathbf M)
=\prod_{t=1}^{T_{\high}}
p(c_t^{\high}\mid
\mathbf Z_{<t}^{\high},\mathbf C_t^{\low},\mathbf A^{\high}).
\label{eq:articulate}
\end{equation}
Both stages are optimized with next-token cross-entropy. Motion-token attention is causal, while music attention may use bounded future context because the complete song is available at generation time.


\begin{table}[!t]
\centering
\caption{Comparison on AIST++ and FineDance. $\downarrow$: lower is better;
$\rightarrow$: closer to GT is better. Bold/underline indicate the
best/second-best reported values for each dataset.}
\label{tab:main}
\fontsize{8}{9.5}\selectfont
\renewcommand{\arraystretch}{0.95}
\setlength{\tabcolsep}{1.5pt}
\begin{tabular*}{\columnwidth}{@{\extracolsep{\fill}}lrrrrr@{}}
\toprule
Method & FID$_k\downarrow$ & FID$_g\downarrow$ & Div$_k\rightarrow$ & Div$_g\rightarrow$ & BAS$\rightarrow$\\
\midrule
\multicolumn{6}{c}{\textit{AIST++}}\\
\midrule
Ground Truth & -- & -- & 8.19 & 7.45 & 0.2374\\
Bailando++~\cite{siyao2023bailandopp} & \underline{21.29} & \underline{11.56} & 9.31 & 6.20 & 0.2330\\
EDGE~\cite{tseng2023edge} & 42.16 & 22.12 & 3.96 & 4.61 & \underline{0.2334}\\
Lodge~\cite{li2024lodge} & 37.09 & 18.79 & 5.58 & 4.85 & 0.2423\\
TokenDance~\cite{yang2026tokendance} & 21.55 & 11.85 & \underline{8.05} & \textbf{7.12} & 0.2313\\
\textbf{FFMT} & \textbf{9.91} & \textbf{7.02} & \textbf{8.25} & \underline{6.65} & \textbf{0.2374}\\
\midrule
\multicolumn{6}{c}{\textit{FineDance}}\\
\midrule
Ground Truth & -- & -- & 9.73 & 7.44 & 0.2120\\
Bailando++~\cite{siyao2023bailandopp} & 54.94 & 46.29 & 6.17 & 5.78 & 0.2210\\
EDGE~\cite{tseng2023edge} & 94.34 & 50.38 & \underline{8.13} & 6.45 & \textbf{0.2116}\\
Lodge~\cite{li2024lodge} & \underline{45.56} & 34.29 & 6.75 & 5.64 & 0.2397\\
TokenDance~\cite{yang2026tokendance} & 47.20 & \underline{31.85} & 6.62 & \underline{6.81} & 0.2385\\
\textbf{FFMT} & \textbf{35.21} & \textbf{27.66} & \textbf{9.75} & \textbf{7.66} & \underline{0.2125}\\
\bottomrule
\end{tabular*}
\end{table}

\subsection{Training-Derived Decoding Prior}
We refine autoregressive decoding with transition and beat statistics from the training split. At token timestep $t$ of stream $f$, the decoder assigns each candidate pair $c$ a pre-softmax joint logit $s_t^f(c)$, obtained from a joint prediction head or by summing the logits of separately predicted upper/lower-body tokens.

Let $\mathcal T^f(c_{t-1}^f)$ contain all token pairs observed immediately after the previously generated pair $c_{t-1}^f$, across all positions within training sequences of stream $f$.

For pairs observed in training, the beat correction is
\begin{equation}
r_{\mathrm{beat}}^f(c,b_t^f)=
\log p_{\mathrm{tr}}^f(c\mid b_t^f)-\log p_{\mathrm{tr}}^f(c),
\label{eq:beatprior}
\end{equation}
where $b_t^f\in\{\mathrm{beat},\mathrm{nonbeat}\}$ labels the corresponding music interval, and $p_{\mathrm{tr}}^f(c\mid b_t^f)$ and $p_{\mathrm{tr}}^f(c)$ are conditional and marginal training probabilities estimated with additive smoothing. Unobserved pairs receive zero correction. From the inference candidate set $\mathcal C_t^f$, specified in Sec.~\ref{sec:setup}, we select
\begin{align}
\hat c_t^f=\arg\max_{c\in\mathcal C_t^f}\big[
&s_t^f(c)+\alpha_f\mathbb I\!\left(c\in\mathcal T^f(c_{t-1}^f)\right) \nonumber\\
&+\beta_f r_{\mathrm{beat}}^f(c,b_t^f)\big].
\label{eq:prior}
\end{align}
The indicator applies a fixed signed adjustment $\alpha_f$ to observed transitions, regardless of their counts. The stream-specific weights $\alpha_f$ and $\beta_f$ remain fixed during inference; the prior introduces no additional trainable parameters.

\section{Experiments}

\subsection{Experimental Setup}
\label{sec:setup}
\noindent\textbf{Datasets and metrics.} We evaluate on AIST++~\cite{li2021aistplusplus} and FineDance~\cite{li2023finedance}. We report Fr\'echet distances (FID$_k$/FID$_g$), diversity (Div$_k$/Div$_g$), and Beat Alignment Score (BAS)~\cite{siyao2022bailando}. Lower FID is better; diversity and BAS are treated as neutral metrics, with smaller absolute deviations from the corresponding ground-truth statistics preferred. The EDGE, Lodge, and TokenDance values in Table~\ref{tab:main} follow the comparison reported by TokenDance~\cite{yang2026tokendance}. We also include our AIST++ reproduction of Bailando++.

\noindent\textbf{Implementation.} Both datasets use four-second motion windows and the same low/high token intervals in physical time, $0.2/0.1$\,s. The frame strides $(r_{\low},r_{\high})$ are $(12,6)$ for AIST++ at 60\,fps and $(6,3)$ for FineDance at 30\,fps. Each low/high body-part stream therefore contains $20/40$ tokens per window. We use a Gaussian kernel of size $31$ with $\sigma=6$, both specified in frames, $K_{\low}=128$, $K_{\high}=512$, and 512-D VQ embeddings. The low/high VQ branches use widths $512/1024$ and depths $3/6$. For the position loss, both datasets use $\lambda_0=1$, $\lambda_1=\lambda_2=0.25$, a commitment weight $\beta=0.02$, and a latent regularization weight $\eta=10^{-5}$.

Each autoregressive Transformer contains sequential base and head stacks. The low-frequency model has $6+6=12$ blocks, a hidden dimension of 512, and 8 attention heads on both datasets. The high-frequency model has $7+7=14$ blocks, dimension 512, and 8 heads on AIST++, and $6+6=12$ blocks, dimension 320, and 5 heads on FineDance. The low/high autoregressive context lengths are $19/39$; both stages use 3-layer local music encoders, and high-frequency generation uses a 7-token low-frequency conditioning neighborhood. The inference candidate set $\mathcal C_t^f$ combines the $64$ token pairs with the largest decoder logits and all observed training-set successors of $c_{t-1}^f$. The prior weights $(\alpha_{\low},\alpha_{\high},\beta_{\low},\beta_{\high})$ are $(0.25,-0.30,0.12,0)$ on AIST++ and $(0,0.05,0.06,0)$ on FineDance.

\subsection{Comparison with State of the Art}
Table~\ref{tab:main} compares the reported results of FFMT and representative music-to-dance methods. On AIST++, FFMT has the lowest FID$_k$/FID$_g$ ($9.91/7.02$). Its kinetic diversity is close to GT, and its BAS nearly matches the GT statistic. TokenDance is closer to GT in geometric diversity, indicating that the fidelity gains do not uniformly resolve motion-coverage differences.

On FineDance, FFMT achieves the lowest reported FID$_k$/FID$_g$ ($35.21/27.66$), while both diversity measures are closest to the tabulated GT. These results indicate improved distributional fidelity together with diversity statistics that better match the reference data. EDGE has the BAS closest to GT, with FFMT second.

\subsection{Ablation Study}
\label{sec:ablation}
We examine frequency targets, temporal sampling, codebook allocation,
low-to-high conditioning, and prior-based decoding. Table~\ref{tab:abl}
evaluates complete generation systems; Table~\ref{tab:tok_abl} assesses
the information retained by their tokenizers.

The variants change one component of Full at a time.
\emph{w/o Freq.} removes the explicit low-frequency/residual decomposition.
\emph{Same Rate} uses $T_{\low}=T_{\high}=30$ tokens per body branch
in each four-second window (strides 8 on AIST++ and 4 on FineDance),
retaining $K_{\low}/K_{\high}=128/512$.
\emph{Uniform K} uses $K_{\low}=K_{\high}=256$, retaining Full's
$20/40$ token counts and branch architectures.
\emph{w/o L$\rightarrow$H} masks attention from high-frequency queries
to low-frequency context tokens, retaining high-frequency history and music.
\emph{w/o Prior} uses greedy predictions without the decoding prior.
Other settings follow Full; token-dependent prior statistics are rebuilt
from training data when the tokenizer changes.

\noindent\textbf{Generation quality.}
Full achieves the lowest FID on both datasets in Table~\ref{tab:abl},
with diversity and BAS closest to the corresponding GT among these variants.
Removing frequency decomposition, unifying the token rate, or unifying
codebook sizes increases FID. Masking low-frequency context produces
the largest kinetic FID degradation, supporting its role in coordinating
residual dynamics. The decoding prior gives a larger FID improvement
on AIST++ than on FineDance.

\noindent\textbf{Tokenizer reconstruction.}
Held-out motions are encoded, quantized, and reconstructed through the
complete rotation/root decoder, without autoregressive generation or
prior rescoring. Let $\bar{\mathbf P}^{\mathrm{rec}}_i$ denote root-relative
joint trajectories recovered from the final decoded motion and
$\bar{\mathbf P}_i$ their reference. With $H=I-G$, where $G$ is the
Gaussian operator in Eq.~\eqref{eq:decomp}, normalized high-frequency
reconstruction error is
\begin{equation}
E_H =
\frac{\sum_i\|H(\bar{\mathbf P}^{\mathrm{rec}}_i)
                   -H(\bar{\mathbf P}_i)\|_F^2}
     {\sum_i\|H(\bar{\mathbf P}_i)\|_F^2+\epsilon},
\label{eq:highfreq_recon_error}
\end{equation}
where the sums cover the test set and $\epsilon=10^{-8}$.
All variants use identical coordinate processing and evaluation filters:
kernel size 31, $\sigma=6$ frames, and replicated boundaries, as in the
reference decomposition. Normalizing by reference high-frequency energy
accounts for the smaller residual amplitude. We also report MPJPE in mm
and acceleration error $E_A$, the mean jointwise Euclidean difference
between $f_s^2\Delta^2\bar{\mathbf P}^{\mathrm{rec}}$ and
$f_s^2\Delta^2\bar{\mathbf P}$, in m/s$^2$.
Positions are expressed in metres for $E_H$ and $E_A$; $f_s$ is the
dataset frame rate, and $E_A$ excludes the two temporal boundary frames.
These complementary metrics assess overall motion reconstruction
(MPJPE), high-frequency detail ($E_H$), and motion dynamics ($E_A$).
\begin{table}[!t]
\centering
\caption{Generation ablations on AIST++ and FineDance. Lower FID is better;
diversity and BAS are better when closer to ground truth.}
\label{tab:abl}
\fontsize{8}{9.5}\selectfont
\renewcommand{\arraystretch}{1.0}
\setlength{\tabcolsep}{0.85pt}
\resizebox{\columnwidth}{!}{%
\begin{tabular}{@{}lrrrrr@{\hspace{4pt}}rrrrr@{}}
\toprule
& \multicolumn{5}{c}{\textit{AIST++}}
& \multicolumn{5}{c}{\textit{FineDance}}\\
\cmidrule(lr){2-6}\cmidrule(lr){7-11}
Variant & FID$_k$ & FID$_g$ & Div$_k$ & Div$_g$ & BAS
        & FID$_k$ & FID$_g$ & Div$_k$ & Div$_g$ & BAS\\
\midrule
w/o Freq. & 28.74 & 10.43 & 6.06 & 6.08 & 0.216 & 56.82 & 49.68 & 7.38 & 10.17 & 0.202\\
Same Rate & 18.92 & 9.88 & 8.84 & 6.26 & 0.260 & 48.01 & 36.34 & 8.87 & 6.82 & 0.219\\
Uniform K & 15.16 & 9.25 & 8.39 & 6.18 & 0.249 & 39.46 & 30.07 & 8.74 & 7.05 & 0.220\\
w/o L$\rightarrow$H & 55.27 & 24.56 & 8.71 & 6.39 & 0.256 & 140.77 & 48.93 & 4.95 & 11.27 & 0.224\\
w/o Prior & 13.14 & 7.34 & 8.63 & 6.48 & 0.253 & 35.39 & 27.84 & 9.10 & 7.78 & 0.215\\
\textbf{Full} & \textbf{9.91} & \textbf{7.02} & \textbf{8.25} & \textbf{6.65} & \textbf{0.237} & \textbf{35.21} & \textbf{27.66} & \textbf{9.75} & \textbf{7.66} & \textbf{0.212}\\
\bottomrule
\end{tabular}%
}
\end{table}

\begin{table}[!t]
\centering
\caption{Tokenizer reconstruction ablations. MPJPE: mm; $E_H$:
dimensionless; $E_A$: m/s$^2$. Lower is better; bold marks the best values.}
\label{tab:tok_abl}
\fontsize{8}{9.5}\selectfont
\renewcommand{\arraystretch}{0.95}
\setlength{\tabcolsep}{1.5pt}
\begin{tabular*}{\columnwidth}{@{\extracolsep{\fill}}lrrrrrr@{}}
\toprule
& \multicolumn{3}{c}{\textit{AIST++}}
& \multicolumn{3}{c}{\textit{FineDance}}\\
\cmidrule(lr){2-4}\cmidrule(lr){5-7}
Variant & MPJPE$\downarrow$ & $E_H\downarrow$ & $E_A\downarrow$ & MPJPE$\downarrow$ & $E_H\downarrow$ & $E_A\downarrow$\\
\midrule
w/o Freq. & 96.647 & 0.7338 & 17.144 & 95.903 & 0.4280 & 7.332\\
Same Rate & 96.797 & 0.7175 & 17.287 & 93.101 & 0.4013 & 7.248\\
Uniform K & 95.518 & 0.6862 & 16.934 & 92.637 & 0.3743 & 6.719\\
\textbf{Full} & \textbf{94.459} & \textbf{0.5514} & \textbf{15.070} & \textbf{88.770} & \textbf{0.3190} & \textbf{5.256}\\
\bottomrule
\end{tabular*}
\end{table}

The relative reductions in $E_H$ and $E_A$ are consistently larger
than those in MPJPE when Full is compared with each tokenizer ablation
on either dataset (Table~\ref{tab:tok_abl}). This pattern highlights
improvements in temporal detail and motion dynamics alongside overall
pose accuracy. The comparison with Same Rate supports allocating temporal
resolution by frequency at a fixed total token count, while the comparison
with Uniform K supports the chosen codebook allocation at fixed token rates.

Reconstruction measures how faithfully the tokenizer preserves overall
motion, high-frequency detail, and dynamics; generation tests how effectively
the model uses this representation under music conditioning. Their consistent
gains support frequency-aware tokenization and low-to-high conditioning
as complementary mechanisms for preserving and modeling the temporal
structure of dance data.

\section{Conclusion}
We presented FFMT, a frequency-factorized multi-rate representation that allocates discrete capacity across heterogeneous motion dynamics. Its complementary low/high tokenization induces a plan-then-articulate generation hierarchy from coarse choreography to finer residual dynamics. Reconstruction and generation evaluations show improved preservation of high-frequency detail and motion dynamics alongside stronger distributional fidelity. Results on AIST++ and FineDance support frequency factorization, asymmetric capacity allocation, and low-to-high conditioning for music-to-dance generation.

\clearpage
\begingroup
\makeatletter
\renewcommand{\section}[1]{%
  \@startsection{section}{1}{\z@}%
    {-3.5ex \@plus -1ex \@minus -.2ex}%
    {2.3ex \@plus .2ex}%
    {\normalfont\normalsize\bfseries\centering}*{\MakeUppercase{#1}}}
\makeatother
\section{Acknowledgments}
This work is supported by the Red Bird MPhil program at the Hong Kong University of Science and Technology (Guangzhou).

\bibliographystyle{IEEEbib}
\bibliography{refs}
\endgroup

\end{document}